\documentclass[12pt,preprint]{aastex}
\usepackage{natbib}
\usepackage{graphicx}
\usepackage{amssymb}
\usepackage{amsmath}
\shorttitle{}
\begin{document}


\title{Indirect Evidence for Dark Matter Density Spikes around Stellar-Mass Black Holes}


\author{Man Ho Chan, Chak Man Lee}
\affil{Department of Science and Environmental Studies, The Education University of Hong Kong \\
Tai Po, New Territories, Hong Kong, China}


\begin{abstract}
It has been suggested for a long time that dark matter would form a density spike around a black hole. However, no promising evidence has been observed so far to verify this theoretical suggestion. Here, we report the existence of a dark matter density spike around each of the two nearby stellar-mass black holes (A0620-00 and XTE J1118+480). The dynamical friction between dark matter and the companion stars can satisfactorily explain the abnormally fast orbital decays in the two binaries. The calculated spike index for A0620-00 and XTE J1118+480 are $\gamma=1.71^{+0.01}_{-0.02}$ and $\gamma=1.85^{+0.04}_{-0.04}$ respectively, which are close to the lower regime predicted by the stellar heating model. It may provide a possible indirect evidence for the existence of dark matter density spikes around stellar-mass black holes. We anticipate that analyzing observational data of nearby black hole X-ray binaries would be a new way to reveal the nature of dark matter.
\end{abstract}

\section{Introduction}
Theoretical calculations have predicted that dark matter density distribution would be altered by a massive black hole \citep{Gondolo,Merritt2,Gnedin,Merritt,Sadeghian,Nampalliwar}. The conservation of angular momentum and energy would naturally force dark matter to form a dense spike (i.e. a cusp-like density profile) \citep{Gondolo,Merritt2,Gnedin,Sadeghian}. Generally speaking, dark matter density around a black hole would eventually follow a simple power-law form: $\rho_{\rm DM} \propto r^{-\gamma}$, where $r$ is the radial distance from the black hole and $\gamma$ is the spike index. The value of $\gamma$ is model-dependent, which can range from $\gamma=1.5$ to $\gamma=2.5$ \citep{Merritt2,Gnedin,Sadeghian,Fields,Lacroix}. Since the dark matter density profile is a singular form in $r$, the dark matter density near black hole would be very high (i.e. a dense spike).

Based on this theoretical prediction, if dark matter can self-annihilate to give gamma-ray photons, one can expect that the annihilation gamma-ray signals would be greatly enhanced because the annihilation rate is proportional to $\rho_{\rm DM}^2$. A lot of attention has been paid specifically for the dark matter density spike surrounding galactic supermassive black holes \citep{Gondolo,Gnedin,Fields,Bertone,Shapiro} and intermediate-mass black holes \citep{Lacroix,Chan}. Various studies have been done to examine the possible enhanced gamma-ray signals, especially near the supermassive black hole in the Milky Way galaxy \citep{Fields,Shapiro}. However, no promising signals have been observed to verify the theoretical prediction \citep{Fields}. However, it does not mean that the dark matter density spike model is wrong. The negative result in gamma-ray observations could be due to the following reasons: 1. the rest mass of dark matter particles is very large, 2. the annihilation cross section is very small, or 3. dark matter particles do not self-annihilate.

On the other hand, observations and studies of the two closest black hole low-mass X-ray binaries (BH-LMXBs), A0620-00 and XTE J1118+480, have provided very precise measurements for many important physical parameters, including the orbital period $P$, observed radial velocity of the companion star $K$, orbital inclination $i$, black hole mass $M_{\rm BH}$, and the mass of the companion star $m$ (or the mass ratio $q=m/M_{\rm BH}$) \citep{McClintock,Neilsen,Cantrell,Grunsven,Khargharia,Zurita,Cherepashchuk} (see Table 1 for the measured values). The companion star is orbiting the black hole in a nearly circular orbit for each binary. In particular, observations have revealed abnormally fast orbital decays in the two BH-LMXBs: $\dot{P}=-0.60 \pm 0.08$ ms yr$^{-1}$ for A0620-00 and $\dot{P}=-1.90 \pm 0.57$ ms yr$^{-1}$ for XTE J1118+480 \citep{Gonzalez}. These decays are two orders of magnitude larger than the one expected with gravitational wave radiation \citep{Chen,Chen2}. Standard theories only predict $\dot{P} \sim -0.02$ ms yr$^{-1}$ \citep{Gonzalez3}. Two major proposals have been suggested recently to account for the fast orbital decay. The first one is related to the magnetic braking of the companion star. If the surface magnetic field of the companion star is very strong (e.g. $\ge 10^4$ G), the coupling between the magnetic field and the winds from the companion star driven by X-ray irradiation from the black hole would decrease the orbital period through tidal torques \citep{Chen,Justham}. However, this model requires a significant mass loss from the binary system, which has not been observed \citep{Gonzalez}. The second proposal suggests that the tidal torque between the circumbinary disk and the binary can efficiently extract the orbital angular momentum from the binary to cause the orbital decay \citep{Chen}. Nevertheless, simulations show that the predicted mass transfer rate and the circumbinary disk mass are much greater than the inferred values from observations \citep{Chen}. Although a few recent studies suggesting the resonant interaction between the binary and a surrounding circumbinary disk could produce the observed orbital period decays \citep{Chen2,Xu}, the calculated initial mass and effective temperature of the companion stars somewhat do not match the observations \citep{Chen2}. Therefore, it is still a mystery for the abnormally fast orbital decays in the two BH-LMXBs.

Beside the annihilation rate, the dynamical friction due to dark matter density spike would also be very large. If a star is moving inside a collisionless dark matter background, the star would exert a gravitational force to pull the dark matter particles towards it. Then a concentration of the dark matter particles would locate behind the star and exert a collective gravitational force on the star. This collective gravitational force would slow down the star and the resulting effect is called dynamical friction. The idea of dynamical friction was proposed by Chandrasekhar more than 70 years ago \citep{Chandrasekhar}. However, very surprisingly, the dynamical frictional effect in BH-LMXBs has not been seriously examined in previous studies. Most of the related studies are focusing on the compact binary systems \citep{Antonini,Eda,Pani,Yue,Dai,Li,Becker,Speeney,Kavanagh}. Also, no previous study has realized the possible observable consequence of dark matter density spike surrounding a stellar-mass black hole. In this letter, we will discuss the observed fast orbital decays in the two closest BH-LMXBs with the idea of dynamical friction of dark matter density spike.

\section{The dynamical friction model}
Consider a typical BH-LMXB system. The low-mass companion star with mass $m<1M_{\odot}$ is orbiting a central black hole with mass $M_{\rm BH}$ much greater than the stellar mass. The central black hole is almost stationary at the center of the system. If a dark matter density spike is surrounding the central black hole, the companion star would experience the dynamical friction exerted by dark matter. The energy loss due to dynamical friction would decrease the orbital period $P$ of the companion star.

 The energy loss due to dynamical friction is given by \citep{Chandrasekhar,Yue}:
\begin{equation}
\dot{E}=- \frac{4\pi G^2\mu^2 \rho_{\rm DM} \xi(\sigma) \ln \Lambda}{v},
\end{equation}
where $\mu$ is the reduced mass of the BH-LMXB, $\ln \Lambda \approx \ln (\sqrt{M_{\rm BH}/m})$ is the Coulomb Logarithm \citep{Kavanagh}, $v$ is the orbital velocity, and $\xi(\sigma)$ is a numerical factor which depends on the distribution function and the velocity dispersion $\sigma$ of dark matter. If we assume a Maxwell's distribution for dark matter and take $\sigma=200$ km/s, we will have $\xi(\sigma) \sim 0.9$. However, as the information about dark matter is uncertain, we simply assume $\xi(\sigma)=1$. The orbital velocity can be determined by the observed radial velocity $K$ and the orbital inclination $i$: $v=K/\sin i$.

Using the Keplerian relation $P^2=4\pi^2a^3/G(M_{\rm BH}+m)$ with $a$ being the radius of the orbital motion, we can write
\begin{equation}
\frac{\dot{P}}{P}=\frac{3 \dot{a}}{2a}=-\frac{3 \dot{E}}{2E},
\end{equation}
where $E=-GM_{\rm BH}m/2a$ is the total mechanical energy. Therefore, the orbital decay rate can be expressed in terms of the observed parameter set \{ $q$, $K$, $i$, $P$, $M_{\rm BH}$ \} by:
\begin{equation}
\dot{P}=- \frac{12\pi qGP \ln \Lambda}{(1+q)^2(K/\sin i)} \left[\frac{GM_{\rm BH}(1+q)P^2}{4\pi^2} \right]^{1/3} \rho_{\rm DM},
\end{equation}
where $q=m/M_{\rm BH}$ is the mass ratio.

Following the dark matter density spike theory, dark matter would re-distribute to form a density spike around the black hole in the BH-LMXB within the spike radius $r_{\rm sp}$. We follow the standard assumption $r_{\rm sp}=0.2r_{\rm in}$ used in many other studies \citep{Fields,Eda}, where $r_{\rm in}$ is the radius of black hole's sphere of influence. Outside $r_{\rm sp}$, the dark matter density would follow the local dark matter density of their respective positions in the Milky Way. The dark matter density around the black hole with mass $M_{\rm BH}$ can be modeled by the following profile \citep{Lacroix}:
\begin{equation}
\rho_{\rm DM}=\left\{
\begin{array}{ll}
0 & {\rm for }\,\,\, r\le 2R_s \\
\rho_0 \left(\frac{r}{r_{\rm sp}} \right)^{-\gamma} & {\rm for }\,\,\, 2R_s <r \le r_{\rm sp}, \\
\rho_0 & {\rm for}\,\,\, r>r_{\rm sp} \\
\end{array}
\right.
\end{equation}
where $R_s=2GM_{\rm BH}/c^2$, and $\rho_0$ is the local dark matter density. When the distance from the black hole is larger than the spike radius $r_{\rm sp}$, we assume that the dark matter density would follow back to the local dark matter density. By taking the reference value at the solar position $\rho_{\odot}=0.33 \pm 0.03$ GeV cm$^{-3}$ \citep{Ablimit} and following the Navarro-Frenk-White dark matter density profile \citep{Navarro}, the local dark matter densities of A0620-00 and XTE J1118+480 can be determined by their respective positions \citep{Gonzalez2}: $\rho_0=0.29 \pm 0.03$ GeV cm$^{-3}$ for A0620-00 and $\rho_0=0.34 \pm 0.03$ GeV cm$^{-3}$ for XTE J1118+480.

The radius of influence can be determined by \citep{Merritt,Merritt2}:
\begin{equation}
M_{\rm DM}(r\le r_{\rm in})=\int_0^{r_{\rm in}}4 \pi r^2\rho_{\rm DM}dr=2M_{\rm BH}.
\end{equation}
Therefore, the spike radius $r_{\rm sp}$ is also a function of $M_{\rm BH}$.  Note that the spike density profile assumed here is not an ad hoc parametrization, but follows from theoretical calculations \citep{Gondolo,Sadeghian}. It is mainly determined by the black hole mass.

The spike index $\gamma$ is the only free parameter in this analysis. For a spike of collisionless dark matter that forms about an adiabatically growing black hole, we have $\gamma=2.25-2.5$ \citep{Gondolo,Fields}. However, if gravitational scattering of stars is important, the stellar heating effect would drive the value of $\gamma$ down to a minimum value $\gamma=1.5$ \citep{Merritt2,Gnedin}. Such a change in the spike index depends on the heating time scale, which is given by \citep{Merritt2}:
\begin{eqnarray}
t_{\rm heat}&=&\frac{\sqrt{3\pi} \Gamma(0.5)M_{\rm BH}}{18m \ln \Lambda} \left(\frac{GM_{\rm BH}}{r_{\rm in}^3} \right)^{-1/2}=1.2 \times 10^{15}~{\rm s} \nonumber\\
&&\times \left(\frac{M_{\rm BH}}{5M_{\odot}} \right)^{1/2}\left(\frac{r_{\rm in}}{5~\rm pc} \right)^{3/2} \left(\frac{m}{M_{\odot}} \right)^{-1} \left(\frac{\ln \Lambda}{3} \right)^{-1},
\end{eqnarray}
Here, a constant stellar density and an initial dark matter spike index $\gamma=2.5$ are assumed \citep{Merritt2}. Generally speaking, for the age of the black hole $t_{\rm BH} \ge t_{\rm heat}$, the spike index would approach the minimum value $\gamma=1.5$ more likely.

\section{Results}
\subsection{Constraints on the spike index}
The analytic formula gives $\dot{P}$ in terms of the precisely measured parameters \{ $q$, $K$, $i$, $P$, $M_{\rm BH}$ \}. We find that the typical values of these parameters \{ $0.05$, $500$ km/s, $45^{\circ}$, $1$ day, $5M_{\odot}$ \} in BH-LMXBs can give $\dot{P} \sim -1$ ms yr$^{-1}$ for a typical dark matter spike density $\rho_{\rm DM} \sim 10^{-13}$ g cm$^{-3}$. For our two target BH-LMXBs, we put the corresponding measured parameters and the observed orbital decay rates to constrain the dark matter densities at the respective companion stellar orbits (with radius $a$): $\rho_{\rm DM}(a) \approx 7.65^{+1.62}_{-1.43} \times 10^{-13}$ g cm$^{-3}$ (A0620-00) and $\rho_{\rm DM}(a) \approx 1.60^{+1.51}_{-0.73} \times 10^{-11}$ g cm$^{-3}$ (XTE J1118+480). Following our dark matter density spike model and involving the uncertainties of the measured parameters, we get $\gamma=1.71^{+0.01}_{-0.02}$ for A0620-00 and $\gamma=1.85^{+0.04}_{-0.04}$ for XTE J1118+480 (see Fig.~1 for the general relation between $\dot{P}$ and $\gamma$).

As mentioned above, theoretical predictions give $1.5 \le \gamma \le 2.5$ \citep{Merritt2,Gnedin,Fields,Lacroix}. If the effects of baryons or stellar heating are important, the spike index might be close to the smallest extreme value $\gamma=1.5$ \citep{Merritt2,Gnedin,Fields}. In fact, this stellar heating effect is due to the dynamical friction between stars and dark matter. Therefore, in a BH-LMXB, the continuous gravitational scattering between the companion star and dark matter might provide the similar stellar heating effect to reduce the spike index to a smaller value. Nevertheless, the case for BH-LMXB is somewhat different from the stellar heating scenario discussed in \citet{Merritt2,Gnedin}. There is only one companion object in a BH-LMXB while many stars are involved in the stellar heating scenario. However, recent simulations show that the dynamical friction of the companion object with a large mass ratio $q$ would increase the kinetic energy of the dark matter particles in the halo and somewhat decrease the dark matter density \citep{Kavanagh}, which apparently reduces the spike index. Although this is not identical to the stellar heating scenario, both processes involve the dynamical friction to re-distribute the dark matter density.

 Using the stellar heating scenario as an analogy, we expect that the spike index might be smaller if $t_{\rm BH} \ge t_{\rm heat}$. Using Eq.~(6), the heating time scales for A0620-00 and XTE J1118+480 are $t_{\rm heat}=3.5\times 10^{15}$ s and $t_{\rm heat}=6.1\times 10^{15}$ s respectively. Although we do not know the ages of the black holes in the BH-LMXBs, we can assume $t_{\rm BH} \le P/\dot{P}$. It is because if $t_{\rm BH}>P/\dot{P}$, it should be highly improbable for us to observe A0620-00 and XTE J1118+480 now as both systems would have collapsed very likely within the cosmological age of 13.7 Gyr ($4.3\times 10^{17}$ s), unless a significant change of mass transfer rate has occurred. If we follow this assumption, we can find that the upper limit of $t_{\rm BH}$ for the A0620-00 black hole is the same order of magnitude as the heating time scale while the upper limit of $t_{\rm BH}$ for the XTE J1118+480 black hole is about 20 times smaller than the heating time scale (see Table 2). This may explain why the spike index for A0620-00 is smaller. Therefore, our results reveal a consistent picture for the dark matter spike model and provide a very good explanation for the abnormally fast orbital decay in the two closest BH-LMXBs. Note that our major conclusion still holds even if the stellar heating scenario is not a good analogy.

\subsection{The effect of dark matter annihilation}
We did not assume any dark matter annihilation in the above discussion. If dark matter annihilation rate is large enough, the central dark matter density would approach the constant saturation density $\rho_{\rm sat}=m_{\rm DM}/\langle \sigma v \rangle t_{\rm BH}$ \citep{Lacroix} when $\rho_0(r/r_{\rm in})^{-\gamma}> \rho_{\rm sat}$, where $m_{\rm DM}$ is the mass of a dark matter particle and $\langle \sigma v \rangle$ is the annihilation cross section. If the orbital decays originate from the dynamical friction of dark matter with the saturation density (i.e. the orbital radius is smaller than the saturation radius), we can determine the upper limits of dark matter mass for this particular scenario. Taking the thermal annihilation cross section $\langle \sigma v \rangle=2.2\times 10^{-26}$ cm$^3$/s predicted by standard cosmology \citep{Steigman} and the upper limits of $t_{\rm BH}$, we can get $m_{\rm DM} \le 14$ GeV for A0620-00 and $m_{\rm DM} \le 48$ GeV for XTE J1118+480 if the companion stars are moving in the dark matter saturation density region. In other words, if $m_{\rm DM}>48$ GeV, the companion stars in both systems would be orbiting the corresponding black hole in the dark matter density spike region. Since many recent stringent constraints of thermal annihilating dark matter indicate $m_{\rm DM} \ge 100$ GeV \citep{Ackermann,Chan2,Abazajian,Regis}, the dark matter density would not be saturated at the orbital positions in A0620-00 and XTE J1118+480.

\section{Discussion}
The existence of dark matter density spike surrounding a black hole has been suggested for more than two decades. However, no smoking-gun evidence has been obtained from observations. Here, we show that the effect of dynamical friction due to dark matter density spike can satisfactorily explain the fast orbital decay in the two closest BH-LMXBs.  The resultant spike index is $\gamma=1.7-1.8$, which is close to the value predicted by the stellar heating model ($\gamma=1.5$) \citep{Merritt2,Gnedin}. Although the BH-LMXBs considered here are not identical to the stellar heating scenario discussed in \citet{Gnedin}, recent simulations of the compact-object inspirals show that the motion of the companion object would affect the distribution of the dark matter density spike surrounding an intermediate-mass black hole, especially for the mass ratio $q>10^{-3}$ \citep{Kavanagh}. Therefore, we may also see similar results of the stellar heating effect in the BH-LMXBs. Note that although the dark matter density is changing in time during re-distribution, the dynamical friction expression used in Eq.~(1) is still applicable because the change is very slow in time \citep{Kavanagh}. An overall consistent picture can be described as follows. When the black hole in a BH-LMXB is formed, the surrounding dark matter would be re-distributed to form a density spike (probably with an initial spike index $\gamma \approx 2-2.5$) \citep{Gondolo}. However, the dynamical friction between dark matter and the companion star eventually help re-distribute the dark matter density spike again to reduce the spike index to approach $\gamma=1.7-1.8$. The orbital period is also decreasing with a fast rate $\sim 1$ ms yr$^{-1}$ due to dynamical friction. If the age of the black hole is larger than the heating time scale, the final spike index may change to a smaller value.

We can get very small uncertainties in $\gamma$ because the uncertainties of the measured parameters are very small, especially for A0620-00.  The uncertain factor $\xi(\sigma)$ would only change the resulting spike index slightly. Generally speaking, our results may suggest a possible evidence of the existence of dark matter density spike surrounding a black hole. It also suggests that a dark matter density spike might exist around a stellar-mass black hole ($M_{\rm BH} \sim 1-10M_{\odot}$), but not only around a supermassive black hole \citep{Gondolo,Merritt2,Gnedin,Lacroix2} or an intermediate-mass black hole \citep{Lacroix,Dai,Li} as suggested in the past literature. Since no previous study has focused on the case of dark matter density spike around a stellar-mass black hole, the effect of dark matter dynamical friction has also been neglected. In fact, one recent study has proposed that the electron excess detected by the DAMPE experiment might originate from the annihilating dark matter density spike in A0620-00 \citep{Chan3}. Therefore, analyzing the effect of dark matter dynamical friction in BH-LMXBs would open a new independent way for investigating the dark matter distribution near stellar-mass black holes.

Moreover, if dark matter annihilation effect is important so that the central dark matter density becomes saturated, we can calculate the upper limits of dark matter mass for this particular scenario. Since the calculated upper limits of thermal annihilating dark matter mass $m_{\rm DM}$ are generally smaller than the lower limits constrained from recent multi-wavelength studies, the companion stars should be orbiting inside the dark matter density spike rather than the saturation density. In other words, the effect of annihilation is not important in constraining the spike index.

In fact, analyzing the effect of dynamical friction of dark matter density spike in a binary system is not a new idea. Nevertheless, most of the related studies have focused on the binaries of the compact objects (e.g. black hole binaries) rather than the BH-LMXB systems \citep{Eda,Pani,Yue,Dai,Li,Becker,Speeney,Kavanagh}. In compact binaries, both gravitational radiation and dynamical friction of dark matter are significant. Therefore, gravitational wave detection might be required to reveal the nature of dark matter, which might contribute extra uncertainties in the constrained parameters. Since optical and X-ray observations can give very precise measurements for most of the important physical parameters in BH-LMXBs, we anticipate that analyzing BH-LMXBs can better reveal the nature of the dark matter density spike surrounding a black hole. There are at least 18 black hole X-ray binaries in our Galaxy \citep{Chen}, which can give rich information to constrain the nature of dark matter. For example, one nearby black hole X-ray binary Nova Muscae 1991 also shows an abnormally fast orbital decay $\dot{P}=-20.7 \pm 12.7$ ms yr$^{-1}$, although the uncertainty is quite large \citep{Gonzalez3}. Future high quality measurements may be helpful to further confirm the existence of dark matter density spike in these black hole X-ray binaries. This kind of analyses would open an entirely new window for observations and theoretical studies to investigate dark matter astrophysics \citep{Bertone2}.

\begin{table}
\caption{The measured parameters of A0620-00 and XTE J1118+480.}
\begin{tabular}{ |l|l|l|}
 \hline\hline
        & A0620-00 & XTE J1118+480 \\
 \hline
  $M_{\rm BH}$ & $5.86 \pm 0.24 M_{\odot}$ \citep{Grunsven} & $7.46^{+0.34}_{-0.69}M_{\odot}$ \citep{Gonzalez} \\
  $q$ & $0.060 \pm 0.004$ \citep{Grunsven} & $0.024 \pm 0.009$ \citep{Khargharia} \\
  $K$ (km/s) & $435.4 \pm 0.5$ \citep{Neilsen} & $708.8 \pm 1.4$ \citep{Khargharia} \\
  $i$ & $54^{\circ}.1 \pm 1^{\circ}.1$ \citep{Grunsven} & $73^{\circ}.5 \pm 5^{\circ}.5$ \citep{Khargharia} \\
  $P$ (day) & $0.32301415(7)$ \citep{Gonzalez} & $0.16993404(5)$ \citep{Gonzalez} \\
  $\dot{P}$ (ms yr$^{-1}$) & $-0.60 \pm 0.08$ \citep{Gonzalez} & $-1.90 \pm 0.57$ \citep{Gonzalez} \\
  $d$ (kpc) & $1.06 \pm 0.12$ \citep{Gonzalez2} & $1.70 \pm 0.10$ \citep{Gonzalez2} \\
 \hline\hline
\end{tabular}
\end{table}

\begin{table}
\caption{ The orbital radius $a$, the calculated dark matter density at $a$, the spike index $\gamma$, the radius of influence $r_{\rm in}$, the heating time scale $t_{\rm heat}$, and the upper limit of the black hole age $t_{\rm BH}$ for each BH-LMXB based on the dark matter density spike model.}
\begin{tabular}{ |l|l|l|}
 \hline\hline
         & A0620-00 & XTE J1118+480 \\
\hline
$a$ (AU) & $0.0169^{+0.0003}_{-0.0002}$ & $0.0118^{+0.0002}_{-0.0004}$ \\
$\rho_{\rm DM}(a)$ (g cm$^{-3}$) & $7.65^{+1.62}_{-1.43} \times 10^{-13}$  & $1.60^{+1.51}_{-0.73} \times 10^{-11}$  \\
$\gamma$ &  $1.71^{+0.01}_{-0.02}$ & $1.85^{+0.04}_{-0.04}$ \\
$r_{\rm in}$ (pc) & $5.41^{+0.10}_{-0.09}$ & $5.34^{+0.02}_{-0.06}$ \\
$t_{\rm heat}$ (s) & $3.5 \times 10^{15}$ & $6.1 \times 10^{15}$ \\
$t_{\rm BH}$ (s) & $\le 1.7\times 10^{15}$ & $\le 3.5 \times 10^{14}$ \\
  \hline\hline
\end{tabular}
\end{table}

\begin{figure}
\vskip 10mm
\includegraphics[width=140mm]{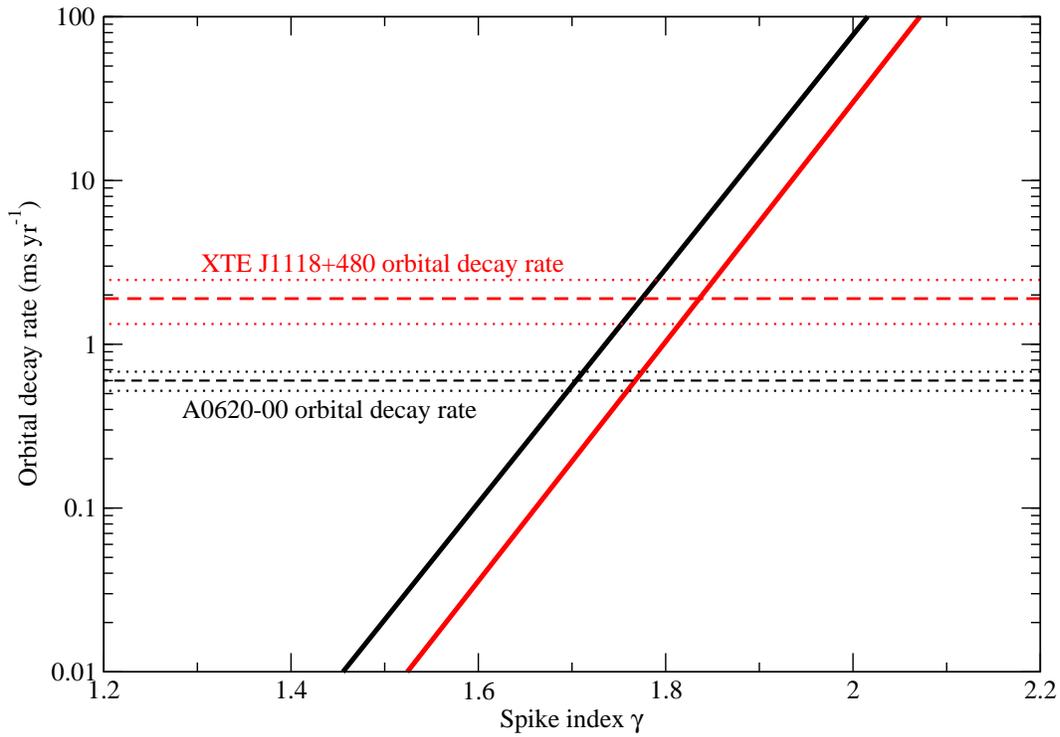}
\caption{The black and red solid lines indicate the relation between $\gamma$ and $\dot{P}$ for A0620-00 and XTE J1118+480 respectively. The horizontal dashed lines and dotted lines represent the mean values and the $1\sigma$ limits of the observed orbital decay rates (black: A0620-00; red: XTE J1118+480).}
\label{Fig1}
\vskip 5mm
\end{figure}

\section{Acknowledgements}
We thank the anonymous referees for useful comments. The work described in this paper was partially supported by a grant from the Research Grants Council of the Hong Kong Special Administrative Region, China (Project No. EdUHK 18300922).


\end{document}